# Universal sheet resistance and revised phase diagram of the cuprate high-temperature superconductors

N. Barišić[a,b,c,d], Y. Li[e], G. Yu[a], X. Zhao[a,f], M. Dressel[b], A. Smontara[c], M. Greven[a]*

[a] School of Physics and Astronomy, University of Minnesota, Minneapolis, Minnesota 55455, USA

[b] 1. Physikalisches Institut, Universität Stuttgart, D-70550 Stuttgart, Germany

[c] Institute of Physics, Bijenička c. 46, HR–10000 Zagreb, Croatia

[d] Service de Physique de l'Etat Condensé, CEA-DSM-IRAMIS, F 91198 Gif-sur-Yvette, France

[e] International Center for Quantum Materials, School of Physics, Peking University, Beijing 100871, China

[f] State Key Lab of Inorganic Synthesis and Preparative Chemistry, College of Chemistry, Jilin University, Changchun 130012, China

* Correspondence to:

Martin Greven
School of Physics and Astronomy, University of Minnesota, Minneapolis, Minnesota 55455, USA
Tel : 612-624-7542
greven@physics.umn.edu

Upon introducing charge carriers into the copper-oxygen sheets of the enigmatic lamellar cuprates the ground state evolves from an insulator into a superconductor, and eventually into a seemingly conventional metal (a Fermi liquid). Much has remained elusive about the nature of this evolution and about the peculiar metallic state at intermediate hole-carrier concentrations ($p$). The planar resistivity of this unconventional metal exhibits a linear temperature dependence ($\rho \propto T$) that is disrupted upon cooling toward the superconducting state by the opening of a partial gap (the pseudogap) on the Fermi surface. Here we first demonstrate for the quintessential compound $HgBa_2CuO_{4+\delta}$ a dramatic switch from linear to purely quadratic (Fermi-liquid-like, $\rho \propto T^2$) resistive behavior in the pseudogap regime. Despite the considerable variation in crystal structures and disorder among different compounds, our result together with prior work gives new insight into the $p$-$T$ phase diagram and reveals the fundamental resistance per copper-oxygen sheet in both linear ($\rho_\Box = A_{1\Box}T$) and quadratic ($\rho_\Box = A_{2\Box}T^2$) regimes, with $A_{1\Box} \propto A_{2\Box} \propto 1/p$. Theoretical models can now be benchmarked against this remarkably simple universal behavior. Deviations from this underlying behavior can be expected to lead to new insights into the non-universal features exhibited by certain compounds.

When exploring the properties of a material, the resistivity is the quantity that is often first measured, but last understood. It is an observable that corresponds to a weighted integration over the whole Fermi surface (see supporting information (SI) Appendix 1). Consequently, the resistivity is highly sensitive to changes in electronic behavior, and hence regularly used to detect phase transitions and determine phase diagrams. Yet for the same reason, the resistivity typically exhibits complex temperature, magnetic field, etc., dependences. Nevertheless, when simple temperature dependences are observed, such as power-law in a metal or exponential in a semiconductor, this usually provides important insight into the fundamental properties of a material[1]. Only in rare cases, such as the quantum Hall effect in correlated two-dimensional electron systems, has the magnitude of the resistivity been found to be the same for an entire class of materials[2].

In the case of the cuprates, the close proximity to the Mott insulator has long been argued to imply a purely electronic superconducting mechanism[3], yet the peculiar metallic state from which the superconductivity emerges upon cooling has remained an enigma. Part of the difficulty stems from the fact that the well over 50 cuprate superconductors exhibit non-universal temperature-doping phase diagrams due to significant differences in superconducting transition temperatures ($T_c$), crystal structures, disorder types, competing phases, etc.[4]

Here we first report quantitative planar dc resistivity results for the structurally simple single-CuO$_2$-layer compound HgBa$_2$CuO$_{4+\delta}$ (Hg1201). Above the pseudogap temperature $T^*$ [5, 6], we observe the well-known yet poorly understood linear behavior $\rho = A_1 T$. Upon further cooling, we find a distinct switch to a purely quadratic dependence, $\rho = A_2 T^2$, that holds between the lower characteristic temperature $T^{**}$ and the onset of superconducting fluctuations at $T'$, just above $T_c$ ($T^* > T^{**} > T' > T_c$). We then present a systematic analysis of prior data for double-layer YBa$_2$Cu$_3$O$_{6+\delta}$ (YBCO)[7,8,9,10,11,12] as well as single-layer La$_{2-x}$Sr$_x$CuO$_4$ (LSCO)[8,9,13,14,15,16] and Tl$_2$Ba$_2$CuO$_{6+\delta}$ (Tl2201)[13,17]. The distinct crystal structures of these four curates are shown in Fig. 1. Due to the quasi-two-dimensional electronic structure of the cuprates, with its underlying square-planar CuO$_2$ sheet, it is natural to consider the resistance $\rho_\square$ per Cu-O sheet (SI Appendix 3), although this is rarely done. Our simple and entirely model-free analysis reveals that, across a large part of the phase diagram, the sheet resistance coefficients $A_{1\square}$ and $A_{2\square}$ are universal and, moreover, proportional to the inverse hole concentration: $A_{1\square} \propto A_{2\square} \propto 1/p$. Our findings give profound new insight into the phase diagram and the nature of the conducting states of the high-$T_c$ cuprates.

**Results**

**Planar resistivity and characteristic temperatures.** Hg1201 crystals were prepared according to a previously reported procedure[18,19]. The in-plane dc resistivities of three underdoped samples with $T_c = 80$ K are displayed in Fig. 2a, and the magnetization characterization for one of them is shown in Fig. 2b. Regardless of their significantly different dimensions (e.g., $a$-$b$ dimensions vary between about 0.01 and 1 mm), the samples exhibit identical temperature dependences (Fig. 2c), and hence a high degree of homogeneity. Extrapolation of the high-temperature linear behavior ($\rho = \rho_0 + A_1 T$) to zero temperature gives $T^* \approx 280$ K and reveals a negligible residual resistivity. Such a small $\rho_0$ is usually taken as the signature of a very clean metallic system, free of extrinsic and intrinsic disorder. This is noteworthy, since the underdoped cuprates typically exhibit a large $\rho_0$ that increases with decreasing doping[6,20].

The quadratic temperature dependence is demonstrated in Fig. 2d. For the $T_c = 80$ K samples, it is found below $T^{**} \approx 170$ K and spans about 80 K in temperature before the influence of superconducting fluctuations is noticeable at $T'$. As shown in Fig. 3a-c, this behavior is also observed at lower doping (samples with $T_c = 47$ K and 67 K). In other cuprates, the pure underlying quadratic behavior tends to be masked: for example, we found no published evidence for $Bi_2Sr_2CuO_{6+\delta}$ (Bi2201) and $Bi_2Sr_2CaCu_2O_{8+\delta}$ (Bi2212), two well-studied compounds for which disorder effects are known to be significant[4] or for other cuprates in the doping range $0.11 < p < 0.30$. However, as summarized in Fig. 3, a quadratic planar resistivity has been observed in a few cases: underdoped YBCO ($p = 0.03$ and $0.09$)[8,10,11], underdoped LSCO ($p = 0.02$ and $0.08$)[8], as well as strongly overdoped LSCO ($p = 0.33$). The latter has been argued to be a Fermi liquid[16]. In Figs. 3e and SI Appendix 5, we demonstrate that $\rho \propto T^2$ also holds at intermediate temperatures for prior data for LSCO at $p = 0.01$[9]. Tl2201[13] ($T_c = 15$ K; Fig. 3i) and LSCO[14] at $p \approx 0.30$ are in close proximity to the putative Fermi-liquid regime, and the description of the planar resistivity requires only a small additional $T$-linear component. For underdoped YBCO at $p \approx 0.11$ in a 55 T magnetic field, an approximately quadratic resistive behavior was reported down to very low temperature[12].

In Fig. 4b and SI Appendix 6 we demonstrate that the four characteristic temperatures and the underlying (hidden) quadratic resistive regimes of YBCO, LSCO and Bi2201 can be identified consistently from prior contour plots of the second temperature derivative of the resistivity[20]. Hg1201 and YBCO are structurally very different, with one and two $CuO_2$ sheets per unit cell, respectively (Fig. 1). However, as shown in Fig. 4, the doping-dependent temperatures $T^*$, $T^{**}$, $T'$ and $T_c$ determined from resistivity, which demarcate five distinct physical regions, are very similar. In both compounds, the opening of the pseudogap at $T^*$ has been shown to be associated with a phase transition to a novel magnetically ordered state[21]. The onset of the $\rho \propto T^2$ behavior below $T^{**}$ agrees surprisingly well with characteristic temperatures determined by two other probes: the peak in the thermoelectric power (TEP) for both Hg1201[22] and YBCO[23,24] (see also SI Appendix 1), and the onset of a Kerr rotation signal for YBCO[25]. In both compounds, superconducting fluctuations affect the dc conductivity only near $T_c$ (below $T'$).

**Universal sheet resistance.** We now analyze the doping dependences of the linear and quadratic contributions to the sheet resistance for Hg1201, YBCO[7,8,10,11,12], Tl2201[13,17] and LSCO[8,9,13,14,15]. As shown in Figs. 4&5, three primary regions need to be distinguished: the $T$-linear regime ($p < p^* \approx 0.19$ and $T > T^*$) and the two seemingly disconnected quadratic regimes ($p < p^*$ and $T < T^{**}$; $p > 0.30$ and $T < 55$ K).

For several reasons, the results in Figs. 2&5 are remarkable. *First*, for underdoped Hg1201, we observe a clear and dramatic 'switch' of scattering mechanisms upon cooling: there is no discernible quadratic (linear) contribution above $T^*$ (between $T^{**}$ and $T'$) and the residual resistivity is tiny. *Second*, $A_{1\square}$ and $A_{2\square}$ are universal, despite substantial differences in crystal structure, disorder, and optimal $T_c$ of the four compounds[4]. Consequently, the states near the Fermi level that contribute to the planar transport are essentially identical, and the underlying fundamental planar resistivity in the normal state of the cuprates is now known. *Third*, $A_{1\square}$ (for $p < p^*$) and $A_{2\square}$ (except near $p^*$) are, to a good approximation, simply proportional to the inverse hole concentration. *Fourth*, the scattering mechanism responsible for the linear temperature dependence of the resistivity is clearly related to fluctuations that disappear upon cooling below $T^*$ and doping beyond $p^*$. This is apparent from the fact that purely $T$-linear behavior is observed only above $T^*$, and also from the behavior of the resistivity just above $p^*$ (Fig. 5b), where $A_{1\square}$ for both LSCO and Tl2201[13,14] decreases faster than $1/p$ and approaches zero as superconductivity disappears around $p = 0.30$.

**Discussion and Conclusions**

**Discussion of doping dependence.** Based on the prior observation of metallic resistive behavior at low hole concentrations, a real-space picture of mesoscopic phase segregation was proposed, with a doping-dependent change of the effective volume relevant to charge transport[9]. However, the evidence for such phase segregation in different cuprate families is varied, which appears difficult to reconcile with our observation of universality over a wide doping range (Fig. 5). Another viewpoint is that much of the cuprate phase diagram is controlled by an underlying quantum critical point[26,27], which is supported by observations of novel magnetism below $T^*$[21,28]. In quantum-critical-point theories, the effective interactions among electrons, and consequently all single-particle renormalization phenomena are assumed to be controlled by a fluctuating order parameter of some kind. Scattering off such fluctuations for $T > T*$ is proposed to cause the linear-$T$ dependence of the resistivity. Interpreted in this fashion, the result in Fig. 5 indicates that the critical fluctuations either condense below $T*$ ($p < p*$) or gradually disappear ($p > p*$). We note though that the $1/p$ dependence of $A_{1\square}$ and the quadratic resistive behavior for $T < T**$ are not predicted by existing quantum-critical-point theories.

A more specific picture is obtained by using the Drude formula $\rho = m*/(ne^2\tau)$, which only assumes that it is possible to separate the scattering rate ($1/\tau$) from the effective density per unit mass ($n/m*$) of carriers[29]. The observation of distinct power-law behaviors ($\rho_\square = A_{1\square}T$ and $A_{2\square}T^2$) over a wide doping range below $p*$ suggests that the scattering rate is proportional to $T$ and $T^2$ in the respective regions of the phase diagram. One interpretation of the result in Fig. 5 is that the doping dependences of the scattering rates and carrier densities in the Drude expression compensate exactly in such a way that $A_{1\square} \propto A_{2\square} \propto 1/p$. However, the simplest interpretation of this proportionality for $p < p*$ is to associate the doping dependence of the resistivity solely with the doped carriers: $n = p$. It follows that the respective scattering rates $\tau_1$ and $\tau_2$ as well as the effective mass $m*$ are doping-independent and universal. This interpretation is consistent with several experimental observations[9,30,31,32], such as the notion that the Fermi arc length in the pseudogap regime is proportional to $p$[31,33], or that the optical effective mass enhancement is doping-independent whereas the effective optical charge density varies almost linearly with doping.[32]

Furthermore, since $A_{1\square} \propto A_{2\square} \propto 1/p$ for $p < p*$, the effective number of carriers would seem to remain unaffected by the closing the pseudogap with increasing temperature. The dominant contribution to the conductivity at all temperatures might therefore come from the relatively fast carriers in the nodal regions[34,35]. Alternatively, to explain the nodal character of the cuprates, one might invoke either the scattering off magnetic fluctuations (hot spots) associated with the antinodes or electronic scattering involving van Hove singularities near the Fermi surface[36]. Finally, the antinodal states might be at the Planckian dissipation limit[14], beyond which coherent single particle propagation is inhibited[37,38]. It has been argued that the pseudogap formation can then be viewed as the lowering of the electronic energy in response to this intense scattering, preventing the scattering from remnant quasiparticle states into the 'hot' antinodal regions[14].

It is instructive to extend the above interpretation based on the Drude formula to the overdoped regime. For $p > p*$, the cuprates feature a large Fermi surface volume, which in accordance with Luttinger's theorem corresponds to $1 + p$ rather than $p$ carriers.[39,40] For hole concentrations between $p*$ and $p \approx 0.30$, the planar resistivity of overdoped LSCO and Tl2201 exhibits both linear and quadratic contributions below $\sim 200$ K in a high $c$-axis magnetic field.[13,14] We note that the linear term smoothly connects to the (zero-field) $A_{1\square} \propto 1/p$ behavior for $p < p*$ and $T > T*$ (Fig. 5). Above $p*$, purely Fermi-liquid-like quadratic resistivity has been observed only for LSCO at $p = x = 0.33$, the highest doping level attained in the cuprates[16] (see also SI Appendix 7). Unlike for $A_{1\square}$, the determination of $A_{2\square}$ near $p*$ is ambiguous, as it depends on the choice of polynomial or parallel resistor fit[14]. However, above $p \approx 0.26$ the contribution of the linear term is small, and the simple relation $A_{2\square} \propto 1/p$ appears to hold again. This suggests that the scattering mechanism in the pseudogap regime might be the same as in the putative Fermi-liquid state at high doping. Clarification of the nodal/antinodal dichotomy, which can be related naturally to the oxygen/copper dichotomy in real space[38] (SI Appendix 8), is essential for understanding this connection.

**Pseudogap phenomenon vs. Fermi liquid.** The underdoped cuprates have been suggested to be 'nodal metals' described by a two-component optical conductivity, with a low-energy Drude component associated with coherent quasiparticles on the Fermi arcs observed in photoemission experiments[10,30]. Photoemission experiments also indicate a nearly material and doping independent near-nodal band structure characterized by a sizable Fermi velocity on the arcs[34,35]. Furthermore, the in-plane infrared spectral weight is found to be insensitive to the opening of the pseudogap[41]. These results are consistent with our observations.

The quadratic resistive behavior in the pseudogap regime extends to rather high temperatures (Fig. 3 and SI Appendix 9) and is cut off at low temperature by either (non-universal) charge localization effects[9,11] or superconducting fluctuations. We emphasize that an approximately quadratic temperature dependence was also reported for YBCO ($p \approx 0.11$) down to low temperature (~ 4 K) (Fig. 4b) after the suppression of superconductivity by a high magnetic field[12], and that the corresponding value of $A_{2\square}$ falls on the universal plots of Figs. 5e&f. As shown in Fig. 2a, for Hg1201 at temperatures below $T**$, the mean-free path is considerably larger than the planar lattice constant, thus the Ioffe-Regel criterion[42] for a good metal is satisfied (SI Appendix 10). The observation for YBCO of quantum oscillations in high magnetic fields appears to provide additional support for a Fermi-liquid state in the underdoped cuprates[43], although this phenomenon occurs in a rather narrow doping range, is suggestive of (small) electron pockets, and has not yet been shown to be universal. Additional evidence for seemingly conventional metallic behavior comes from the well-known $T^2$ dependence of the Hall angle (SI Appendix 6)[8]. Furthermore, motivated by the present work, optical conductivity measurements have revealed a Fermi-liquid-like quadratic frequency dependence and temperature-frequency scaling in Hg1201 samples in which the pure $\rho \propto T^2$ behavior is demonstrated from dc transport[44]. Nevertheless, the situation is highly unconventional, as exemplified by the presence of arcs and by the evidence for a quantum critical point at $p*$. Interestingly, NMR results for LSCO and Hg1201 imply a two-component local magnetic susceptibility in the pseudogap regime, one temperature-independent above $T_c$, i.e., Fermi-liquid-like, and the other temperature-dependent and correlated with the formation of the pseudogap[45]. Further evidence for a two-component quasiparticle spectrum for $p < p*$ comes from a recent analysis of the electronic entropy in several cuprates.[46]

**Updated phase diagram.** Figure 6 shows the updated phase diagram of the cuprates. Figure 6a focuses on the non-superconducting properties, neglecting complications (Fig. 6b) related to low structural symmetry, disorder and localization effects[4,9,11], the 'stripe' instability near $p \approx 1/8$[47], and inhomogeneities near $p*$[14]. The pseudogap temperature $T*$ has been associated with a transition to a state with novel magnetic order in both YBCO and Hg1201[21]. The coincidence of $T**$ with the onset of the polar Kerr-effect in YBCO[25] raises the possibility that this temperature may be universally associated with a second phase transition above $T_c$. There exist at least two qualitatively different scenarios near $p*$ that cannot yet be distinguished. Either both $T*(p)$ and $T**(p)$ approach zero at $p* \approx 0.2$, the putative quantum critical point, or the two characteristic temperatures cross near optimal doping. A possible caveat is that even if much of the phase diagram is controlled by a quantum critical point, where $\rho \propto T$ is expected at all temperatures once superconductivity is suppressed by a sufficiently high magnetic field, this point may be inaccessible through the variation of physical parameters such as doping and magnetic field. Even though the Fermi surface undergoes a change in topology as it breaks up into arcs on the underdoped side of the phase diagram, the electronic scattering mechanism that gives rise to the quadratic behavior appears to be the same at the extreme doping levels of 1% and 33%. In the quadratic resistive regime, the underdoped cuprates are rather good metals and may need to be thought of as a two-component electron fluid as a result of correlations associated with the formation of the pseudogap in which the metallic component evolves from the correlated Fermi-liquid at the highest doping. A $\rho \propto T^2$ behavior was also found on the electron doped side of the phase diagram. However, in contrast to our findings for the hole-doped cuprates, it seems that superconductivity evolves from a state that exhibits $T$-linear resistivity[48] (SI Appendix 6).

The present work demonstrates that the quadratic planar resistivity is a universal property of the underdoped cuprates, and that the structurally simple model compound Hg1201 exhibits negligible residual resistivity and a dramatic switch from $T$-linear to quadratic behavior upon cooling. Although the resistivity is generally highly sensitive to compound-specific characteristics, we have furthermore achieved a quantitative, universal understanding of both the linear and quadratic regimes by considering the resistance per principal building block, the copper-oxygen sheet, of four structurally distinct cuprates. Analysis of compound-specific deviations (e.g., for Bi2201 and Bi2212, or for LSCO near $p = 0.12$) from the underlying behavior reported here can be expected to lead to new insights into the non-universal features exhibited by individual compounds. Most importantly, the present work provides a quantitative basis for the development of a comprehensive theoretical understanding of these fascinating materials.

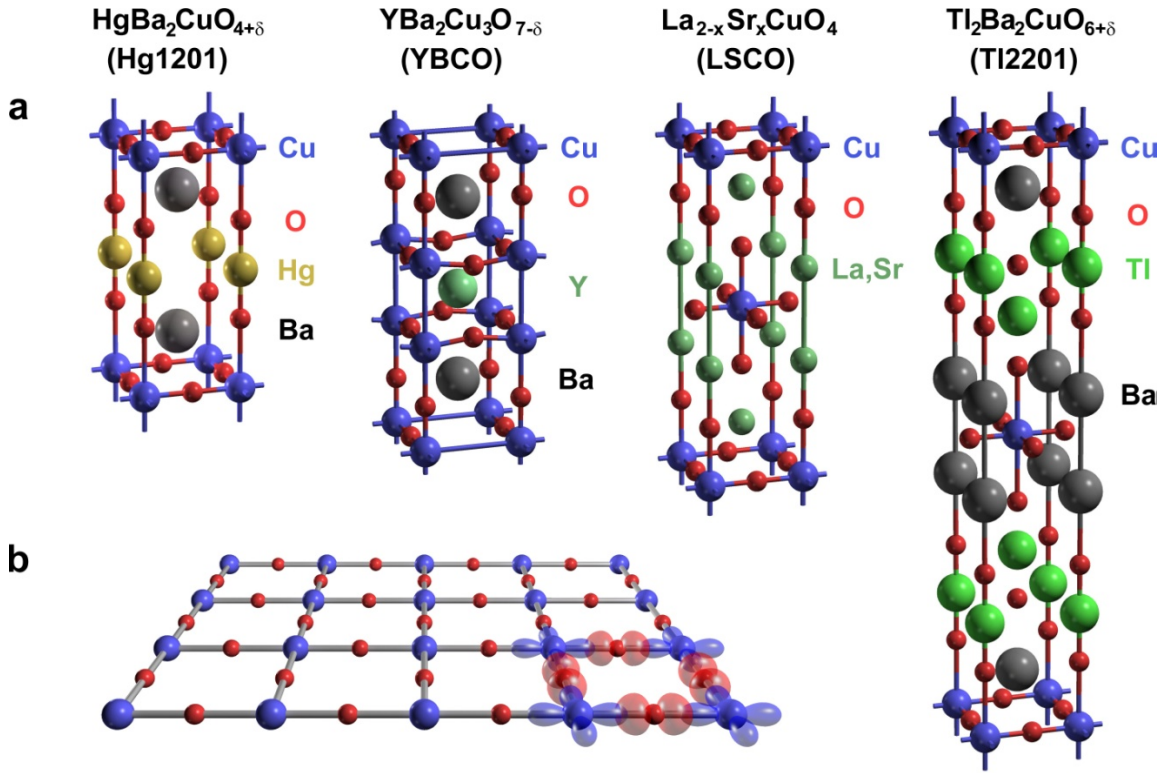

**Fig. 1. Crystal structures of four cuprates.** (a) The unit cells (total number of atoms, individual versus pairs of CuO$_2$ sheets, $c$-axis dimensions, etc.), most prevalent disorder types, and structural symmetry of these four cuprates differ considerably (for details, see Ref. [4] and SI Appendix 2). In Hg1201, YBCO and Tl2201, the hole concentration in the CuO$_2$ sheets is altered by varying the density of interstitial oxygen atoms (each interstitial oxygen introduces up to two holes into nearby CuO$_2$ sheets), whereas in LSCO holes are introduced by replacing La$^{3+}$ with Sr$^{2+}$ ($p=x$ in this case) Hg1201 has a particularly simple crystal structure. It is the first member of the Ruddlesden-Popper family HgBa$_2$CuCa$_{n-1}$Cu$_n$O$_{2n+2+\delta}$, features one CuO$_2$ sheet per formula unit ($n=1$), and the highest optimal $T_c$ ($T_c^{max}$ = 98 K) of all such single-layer compounds (e.g., $T_c^{max}$ = 39 K and 93 K for LSCO and Tl2201, respectively[4]). Furthermore, the physical properties of Hg1201 appear to be least affected by disorder (e.g., the residual resistivity is negligible; see Figs. 2 and 3). (b) The universal building block of the high-$T_c$ cuprates is the CuO$_2$ sheet. The most important electronic orbitals, Cu $d_{x2-y2}$ and O $p_\sigma$, are shown.

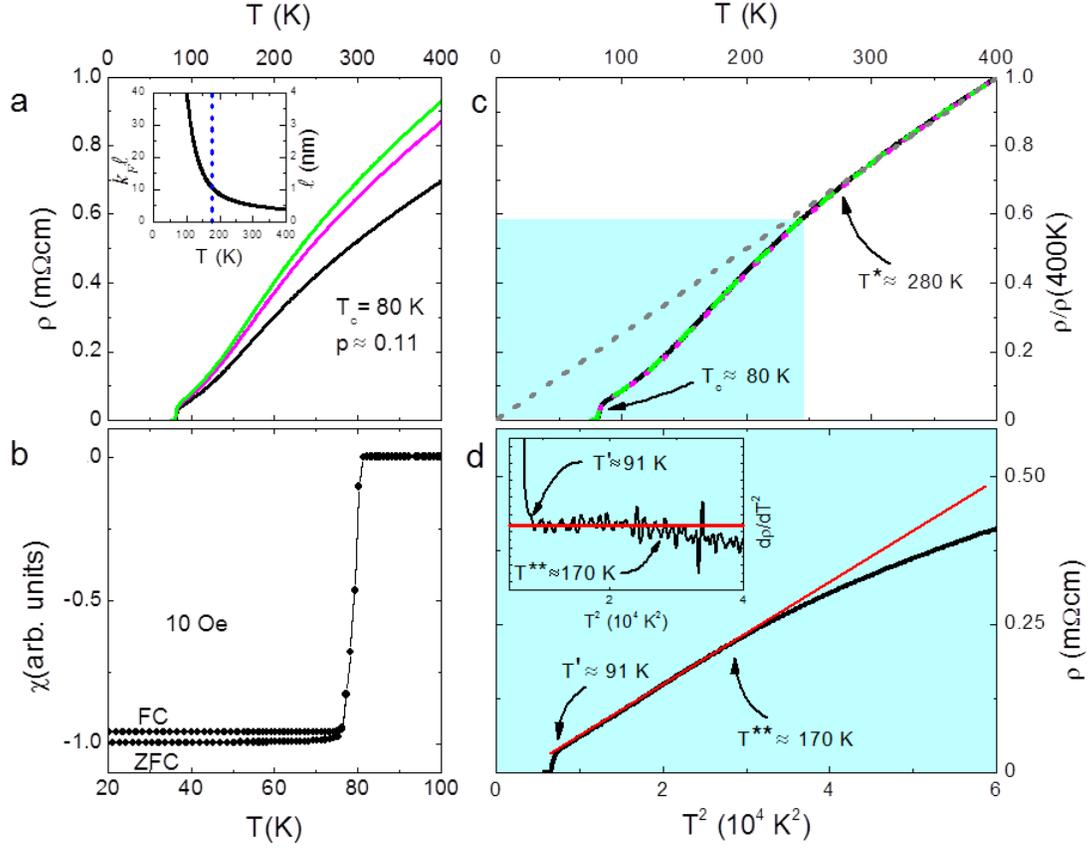

**Fig. 2. *ab*-plane dc-resistivity and magnetic susceptibility for underdoped Hg1201 ($T_c$ = 80 K).** (a) Resistivity as a function of temperature for three samples. Since the samples are cleaved, their shapes are irregular, and consequently the absolute value of the resistivity ($\rho \approx 0.6$ m$\Omega$cm at 300 K) could only be determined with about 20% accuracy (see SI Appendix 4 for experimental methods and detailed results for other hole concentrations). Inset: estimates of mean free path, *l*, and $k_F l$ (SI Appendix 7). Blue vertical dashed line indicates $T^{**}$. (b) Magnetic susceptibility (zero-field-cooled (ZFC) and field-cooled (FC)) for one of the samples after its preparation for resistivity measurements (SI Appendix 4) reveals a sharp onset $T_c$ of 80 K. The FC/ZFC ratio of 97% is exceptionally large, indicative of very high sample quality[19]. (c) When normalized at 400 K, the data in a) collapse onto a single curve, indicating high bulk homogeneity. High-temperature linear behavior, $\rho = \rho_0 + A_1 T$ (dashed line), with $\rho_0 \approx 0$ and $T^* \approx 280$ K. (d) The resistivity exhibits a quadratic temperature dependence between $T' \approx 90$ K and $T^{**} \approx 170$ K. This is also seen from the plot of $d\rho/d(T^2)$ (inset). Red lines are guides to the eye. Horizontal and vertical ranges correspond to blue area in c).

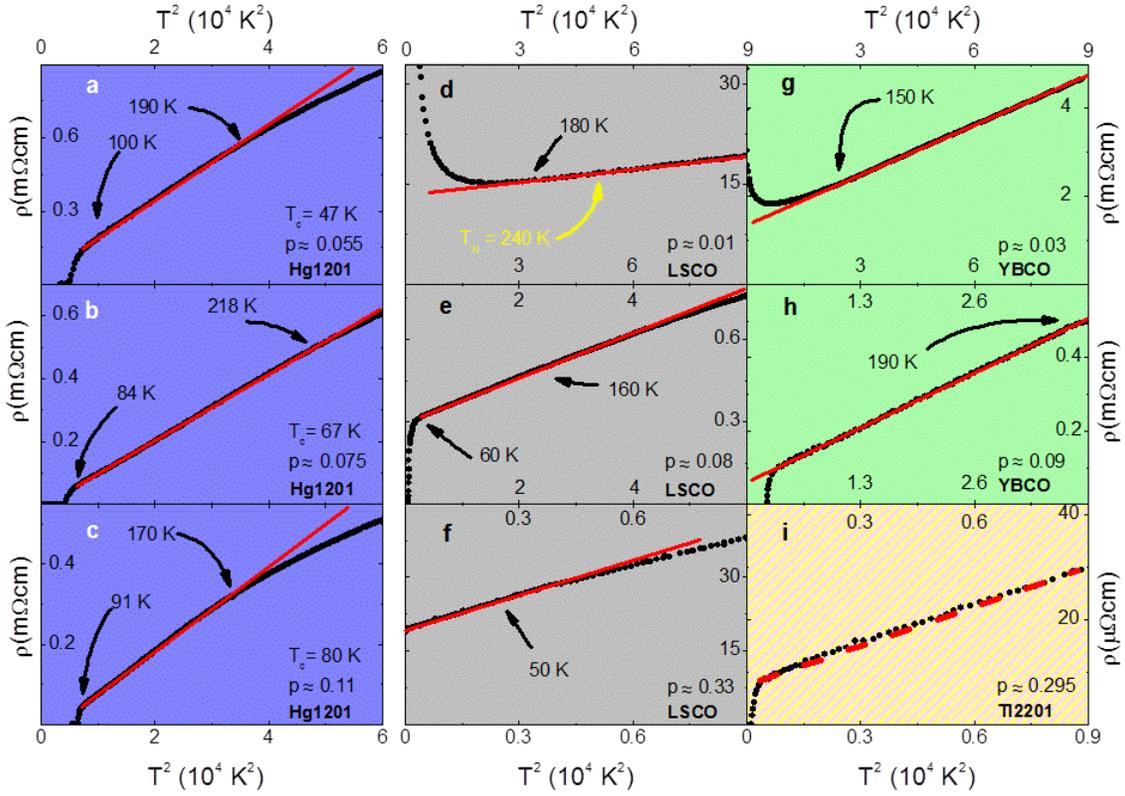

**Fig. 3. Doping and compound dependence of $T^2$-resistivity.** (a-c) Quadratic resistivity in underdoped Hg1201 at three doping levels, observed from $T' \approx T_c + 15$ K (above the superconducting 'tail' for samples with $T_c = 47$ K - see SI Appendix 4) up to $T^{**}$. (d) Similar behavior for LSCO is demonstrated for prior data at $p = 0.01$ (see also SI Appendix 5), where the quadratic behavior persists into the Néel state ($T_N = 240$ K)[9]. For LSCO, $\rho \propto T^2$ was previously reported[8] for $p = 0.02$ (not shown), (e) $p = 0.08$ and (f) $p = 0.33$[16]. For YBCO in zero magnetic field, it was reported for (g) $p = 0.03$[8] and (h) for $p = 0.09$[10]. (i) For Tl2201 at $p = 0.295$, the resistivity increases more slowly than $T^2$ (black dotted curve is slightly concave with respect to the straight dashed red line), which implies the presence of a small linear contribution[13]. Red lines are guides to the eye. Lower ($T_{\rho}'$) and upper ($T_{\rho}^{**}$) temperatures at which resistivity begins to deviate from the quadratic behavior are indicated by arrows. See SI Appendixes 4&6 for estimates of characteristic temperatures and hole concentrations, respectively.

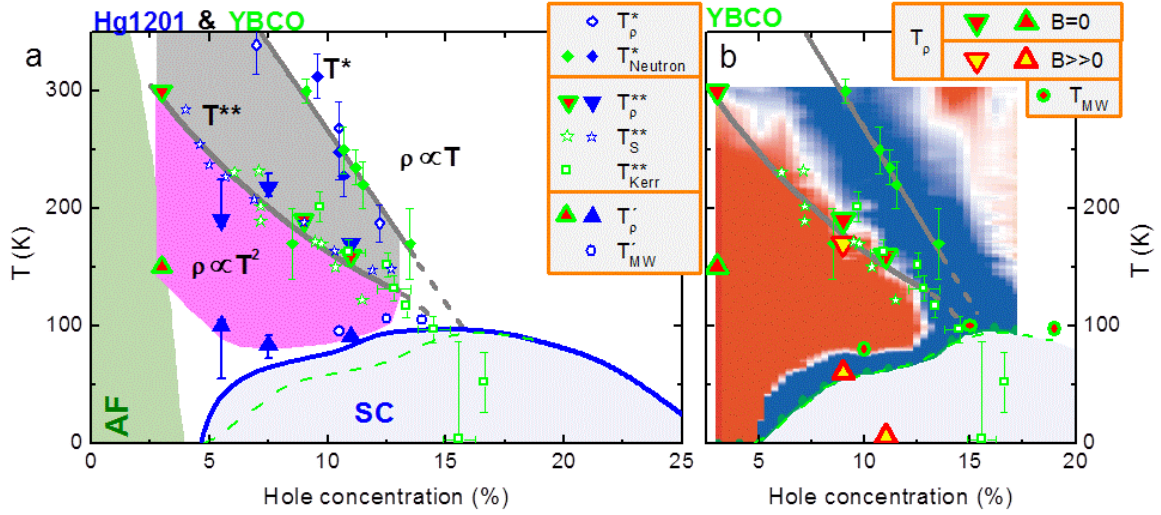

**Fig. 4. Phase diagram of Hg1201 and YBCO.** (a) Pseudogap temperature $T^*$, determined from deviation from linear-$T$ resistivity ($T^*_\rho$) and neutron scattering[21] experiments ($T^*_{Neutron}$). Gray shaded area indicates the crossover to the quadratic regime (magenta) found below $T^{**}$. The latter ends with the onset of SC fluctuations at $T'$, in agreement with microwave ($T'_{MW}$) measurements for Hg1201[49] and YBCO[50], or when localization effects set in (around 150 K for YBCO at $p = 0.03$)[8]. The temperatures of the TEP peak ($T_S^{**}$) for Hg1201[22] and YBCO[23,24] and of the onset of the Kerr effect ($T_{Kerr}^{**}$) for YBCO[25] track $T^{**}_\rho$ from dc-resistivity. Blue and green symbols correspond to Hg1201 and YBCO, respectively. Green dashed line corresponds to the $T_c(p)$ of YBCO. Blue line is obtained from available data for $T_c(p)$ of Hg1201 (up to $p = 0.21$)[22] and extended to higher doping, as discussed in SI Appendix 6. Gray lines for $T^*(p)$ and $T^{**}(p)$ are guides to the eye. Antiferromagnetic (AF) phase is schematically indicated by the green shaded area. (b) The underlying $T^2$ (red contour) regime of YBCO is effectively captured by a map of the resistivity curvature[20]. Quadratic resistive behavior is also apparent after applying a high $c$-axis magnetic field of $B \sim 50$ T[11,12]. For $p = 0.11$[12], the field was sufficiently high to completely suppress the superconductivity and reveal approximately quadratic resistive behavior to low temperature.

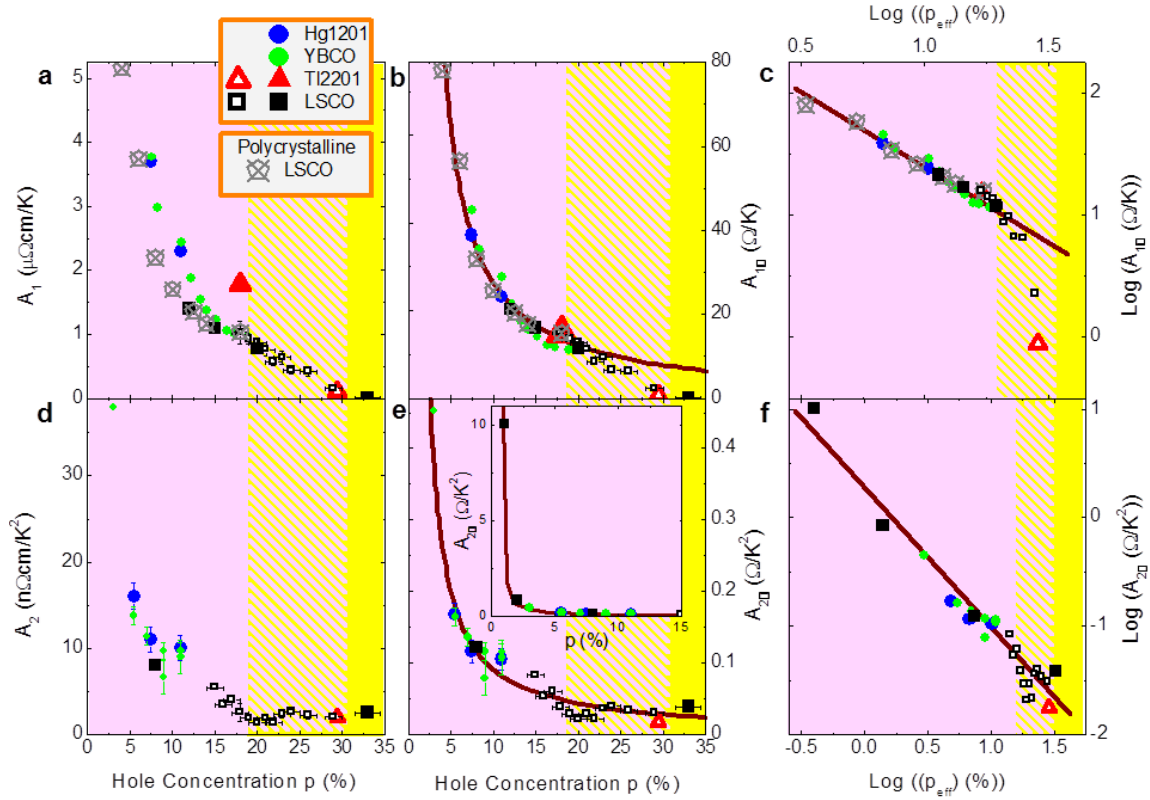

**Fig. 5. Doping dependence of coefficients $A_1$, $A_{1\square}$, $A_2$, and $A_{2\square}$.** There exist three distinct regimes: (i) for $p < 0.19$ (pink), either $\rho \propto A_1 T$ ($T > T^*$) or $\rho \propto A_2 T^2$ ($T^{**} > T > T'$) is observed (filled symbols); (ii) for $0.19 \leq p \leq 0.3$ (pink/yellow hatched area), the resistivity (for LSCO and Tl2201) does not exhibit pure power-law behavior, and prior results from fits to $1/\rho = 1/(\rho_0 + A_1 T + A_2 T^2) + 1/\rho_{max}$, rather than $\rho = \rho_0 + A_1 T + A_2 T^2$, are shown (open symbols); the two forms were found to give closely similar results for $A_1$ at all measured hole concentrations, and for $A_2$ above $p = 0.26$[14]. The latter form yields a lower estimate of $A_2 \approx 0$ near $p^*$; (iii) putative Fermi-liquid regime for $p > 0.3$ (below about 50 K) (yellow). (a,d) $A_1$ and $A_2$ versus hole concentration $p$. (b,e) Demonstration of universal behavior of $A_{1\square}(p)$ and $A_{2\square}(p)$, with approximately constant ratio $A_{1\square}/A_{2\square} \approx 325$ K for $p < 0.19$. (c,f) Log-log plots. A somewhat better linear fit (brown line) is obtained with effective hole concentrations, $A_{1\square}(p_{eff}) \propto 1/(p-p_1)$ (for $0.04 \leq p \leq 0.19$) and $A_{2\square}(p_{eff}) \propto 1/(p-p_2)$ ($0.01 \leq p \leq 0.33$), with nonzero $p_1 = p_2 = 0.007(2)$. For LSCO, data for polycrystalline samples[15] are rescaled by factor of 1/5.5 to match single crystal data. See SI Appendix 6 for estimates of hole concentrations.

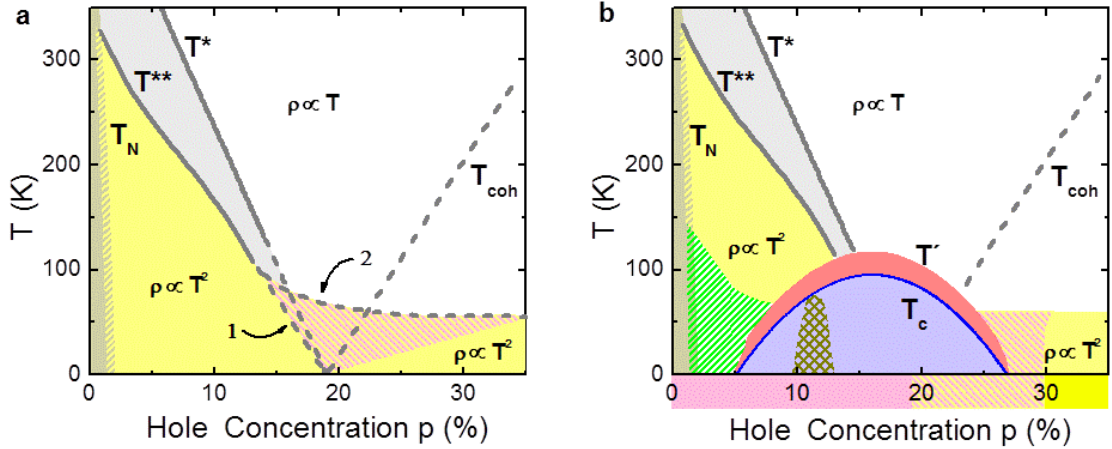

**Fig. 6. Modified phase diagram of the cuprates.** (a) Underlying phase diagram in the absence of disorder and superconductivity. The undoped parent compounds are Mott (charge-transfer) insulators. Antiferromagnetic (AF) order below $T_N$ extends to nonzero doping. Quadratic resistive behavior (observed in LSCO even at $p = 0.01$, for $T > 150$ K) extends into the AF region. The ground state may be insulating up to a small, nonzero hole concentration. $T^*(p)$, and possibly also $T^{**}(p)$ mark phase transitions. $T_{coh}$ corresponds to the loss of antinodal quasiparticle coherence, as observed in photoemission experiments[13]. Two scenarios for $T^{**}(p)$ are indicated by arrows: $T^{**}(p)$ either approaches zero at the putative quantum critical point at $p^* \approx 0.19$ or crosses $T^*(p)$ (hatched area). (b) Phase diagram with disorder, superconductivity and "1/8" anomaly. Superconducting phase: blue; doping/temperature range of the superconducting fluctuations[49,50]: red; localization effects[11]: green; $p = 1/8$ anomaly[47]: olive; possible chemical inhomogeneities[14] (for LSCO) immediately above $p^*$: orange. The three regimes from Fig. 5 are marked along the horizontal axis by corresponding colors.

## Acknowledgements

We gratefully acknowledge discussions with H. Alloul, S. Barišić, A.V. Chubukov, A. Georges, L.P. Gor'kov, S.A. Kivelson, D. van der Marel, F. Rullier-Albenque and A. Shekhter. The work at the University of Minnesota was supported by the Department of Energy, Office of Basic Energy Sciences. The work in Zagreb was supported by the Unity through Knowledge Fund. N.B. acknowledges support by the Alexander von Humboldt Foundation and through a Marie Curie Fellowship.

# Supporting Information Appendix

## Universal sheet resistance and revised phase diagram of the cuprate high-temperature superconductors

N. Barišić, Y. Li, G. Yu, X. Zhao, M. Dressel, A. Smontara, M. Greven

**(1) Resistivity and thermoelectric power**

Starting from the Boltzmann equation, the conductivity $\sigma(T) = 1/\rho(T)$ and the thermoelectric power (TEP) $S(T)$ can be expressed in terms of the Fermi-Dirac distribution $f$, scattering rate $1/\tau$, and electron velocity $v$ (51). For an electric field applied along the $x$ direction, the conductivity is

$$\sigma_x = \frac{e^2}{4\pi^3\hbar} \iint \tau \frac{v_x^2}{v} \, da \, \frac{df}{dE} \, dE, \tag{S1}$$

where the first integral is over a surface of constant energy $E$ and the second integral is over all energies. The Fermi-Dirac distribution in a metal has an appreciable derivative only near the Fermi energy $E_F$, and contributions from states in the $k_B T$ energy window dominate. For example, in the case of a simple cubic crystal, Eq. (S1) simplifies to

$$\sigma = \frac{e^2}{12\pi^3\hbar} \int \tau(k) v(k) \, da \tag{S2}$$

where the velocity and scattering rate depend on the $k$-value of an electron. Not all the electrons near $E_F$ contribute equally. In the cuprates, the antinodal regions of the Brillouin zone do not tend to contribute for electric fields applied along the $CuO_2$ planes.

The TEP is defined as the ratio between the thermal and charge currents,

$$S = \frac{1}{T} \frac{\textit{thermal energy current}}{\textit{charge current}} = \frac{1}{T} \frac{\int_0^\infty \sigma_x(E)(E-E_F)(df/dE)dE}{\int_0^\infty \sigma_x(E)(df/dE)dE} \tag{S3}$$

where $S$ is the Seebeck coefficient and, in accord with Eq. (S1), the partial conductivity is

$$\sigma_x(E) = \frac{e^2}{4\pi^3\hbar} \int \tau \frac{v_x^2}{v} \, da. \tag{S4}$$

For a simple metal, assuming only elastic scattering, this reduces to the Mott expression:

$$S = \frac{\pi^2}{3} \left(\frac{k_B}{e}\right) k_B T \left(\frac{\partial \ln \sigma(E)}{\partial E}\right)_{E=E_F} \tag{S5}$$

It is evident that the TEP is closely related to the (charge) conductivity, and that both are dominated by the band structure in the $k_B T$ energy window around $E_F$. Consequently, both probes are sensitive to changes of the Fermi surface, which will manifest themselves in the temperature dependences of $\rho(T)$ and $S(T)$ as kinks, extrema, etc. Not every such a feature indicates a phase transition. However, when both dc-resistivity and TEP exhibit anomalies over large doping and temperature ranges, this cannot be accidental. Both observables feature striking universalities for different cuprates, indicating that the relevant parts of their Fermi surfaces are very similar.

The TEP of the cuprates can be qualitatively described as follows. At high temperatures, it exhibits linear behavior, $S(T) = S_0 + S_1 T$, with a mysterious (extrapolated) $T = 0$ intercept $S_0$ that becomes dramatically large in underdoped compounds. As the temperature is decreased, $S(T)$ exhibits a peak and a subsequent downturn, and finally falls to zero at $T_c$. Given that $S$ should be zero at $T = 0$ (since the entropy is zero), the eventual downturn upon cooling is expected on general grounds.

The initial deviation from the high-$T$ linear behavior of $S(T)$ appears to be correlated with $T^*(p)$ *(22)* and the maximum of $S(T)$ corresponds to $T^{**}(p)$. Both temperatures are strongly doping-dependent. Neutron diffraction experiments reveal $q = 0$ magnetic order at $T^*$ in YBCO and Hg1201 *(21)*, whereas Kerr-effect measurements give evidence for broken time-reversal symmetry at $T^{**}$ in YBCO *(25)*.

We note that several of our observations are in agreement with the extensive work undertaken by Honma and Hor *(52)*. By analyzing results obtained with numerous experimental probes, they identified three characteristic temperatures above $T_c$, and the lower two correspond to $T^*$ and $T^{**}$ (see also SI Appendix 6).

## (2) Crystal structures

Figure 1 shows the crystal structures of single-layer (*n*=1) Hg1201, LSCO and Tl2201, and of double-layer (*n*=2) YBCO. Hg1201 has space group P4/mmm, with room-temperature lattice constants a ≈ 3.88 Å and c ≈ 9.53 Å *(53)*, and the highest superconducting transition temperature of all *n*=1 cuprates, with $T_c \approx 97$ K at optimal doping. LSCO features a structural transition from tetragonal (space group I4/mmm; a ≈ 3.77 Å and c ≈ 13.22 Å for $p = 0.10$ - $0.20$ *(54)*) at high temperature/hole concentrations to orthorhombic (space group Cmca; e.g., a ≈ 5.35 Å, b ≈ 5.40 Å and c ≈ 13.14 for $p$=0 *(54)*) at low temperature/hole concentration and only $T_c \approx 38$ K at optimal doping. Tl2201 has space group Fmmm, a ≈ 5.47 Å, b ≈ 5.48 Å, c ≈ 23.28 Å and $T_c \approx 93$ K at optimal doping *(55)*. YBCO has space group Pmmm, a ≈ 3.82 Å, b ≈ 3.88 Å, c ≈ 11.65 Å *(56)* and $T_c \approx 93$ K at optimal doping.

## (3) Sheet resistance

Given the quasi-two-dimensional nature of the cuprates, a natural but uncommon way to express the planar resistivity of bulk crystals is as resistivity per Cu-O sheet, the universal building block of the cuprates. The remarkable universal scaling of $A_{1\square}$ and $A_{2\square}$ demonstrated in Fig. 5 for four structurally very different compounds, including both single- and double-layer cuprates, implies that the planar current flow is indeed dominated by the $CuO_2$ planes. For a three-dimensional metal, the resistivity is $\rho \equiv Ra/l$, where $R$ is the measured resistance, $l$ the length along the direction of the current, and *area (a) = width (w) * thickness (t)* the cross-sectional area perpendicular to the current. The sheet resistance $\rho_S = Rw/l = \rho/t$ (which like the resistance $R$ has units of ohm ($\Omega$), although one generally writes $\Omega/\square$ to distinguish the two) is a measure of the resistivity in the case of planar current flow in two-dimensional systems (such as thin films) of uniform thickness. For the cuprates, we consider the resistance per $CuO_2$ plane, $\rho_S = \rho/(c/n)$, where $c$ is the $c$-axis lattice constant and $n$ the number of $CuO_2$ planes per unit cell.

**(4) Experimental details and estimation of characteristic temperatures**

Hg1201 single crystals were grown by a two-step flux method that yields underdoped samples (*18*). Three groups of crystals were subsequently annealed for several weeks under the following conditions: 520 °C in air, 460 °C in 0.1 Torr, and 550 °C at $10^{-6}$ Torr (*19*). This resulted in sharp transitions with $T_c = 80$, 67, and 47 K, respectively. The samples were then cleaved to obtain $ac$-surfaces, on which gold wires (20 μm diameter) were attached with silver paste, subsequently cured for several minutes under the above respective annealing conditions, and then 'quenched' to room temperature. For the $T_c = 80$ K samples, this was a fairly simple task, since it was carried out at ambient pressure. In contrast, curing electrical contacts for the other two sets of samples was delicate, since they needed to be quickly introduced into, and then taken out from the furnace that maintained the low-pressure anneal conditions. Since the samples and the boat on which they were placed required some time to warm up and cool down, equilibrium conditions were never fully reached during the short curing time.

To exemplify the subtleties associated with the curing step, and to explain their consequences on our conclusions, the temperature dependences of the resistivity for four samples annealed to a $T_c$ of 67 K are shown in Fig. S1A. To assure that curing did not change the bulk $T_c$, the magnetic susceptibility was subsequently measured (Fig. S1B). Although after the curing process the bulk transition remained centered at 67 K and sharp (the 10-90% characteristic falls within 1 K), the development of a small 'tail' is now observable toward higher temperatures ($\sim$ 80 K), indicating that a small fraction of the sample has a different (higher) hole concentration (Fig. S1B). This tail is observable in all three measurements: field cooled (FC), zero-field cooled (ZFC), and remnant moment (REM; field cooled, and then measured in zero applied field). The latter is particularly sensitive to inhomogeneities (*19*). The minority contribution affects the resistivity data in a subtle way. Due to filamentary superconductivity, $T_c$ as measured by resistivity appears higher, and the $T^2$ regime is narrowed both from above and below, since contributions from the more highly doped minority regions exhibit both higher $T'$ and lower $T^{**}$. This effect is the reason for the relatively large uncertainty in the extent of the $\rho \propto T^2$ regime for samples at the two lower hole concentrations (Fig. 4). We emphasize that the duration of the curing step is very short compared to the time required to significantly change the bulk oxygen (and hence hole) concentration, so that the vast majority of the sample volume is still at the targeted doping level. Consequently, the resistivity as a bulk probe is not significantly affected in the pure $T$-linear and quadratic regimes, and the coefficients $A_1$ and $A_2$ remain unchanged. We also note that the pseudogap temperature $T^*$, defined as the temperature below which the resistivity deviates from the high-temperature linear behavior, is about 350 K for the $T_c = 67$ K samples, consistent with the characteristic temperature determined from magnetic neutron scattering measurements of the onset of $\boldsymbol{q} = 0$ magnetic order (*21*): the contributions from the more highly doped filamentary regions deviate from linearity at lower temperature than the bulk, and hence do not affect the determination of $T^*$. We conclude that the bulk doping level of our most underdoped samples remains unchanged by the short contact curing, and that the observed properties are intrinsic.

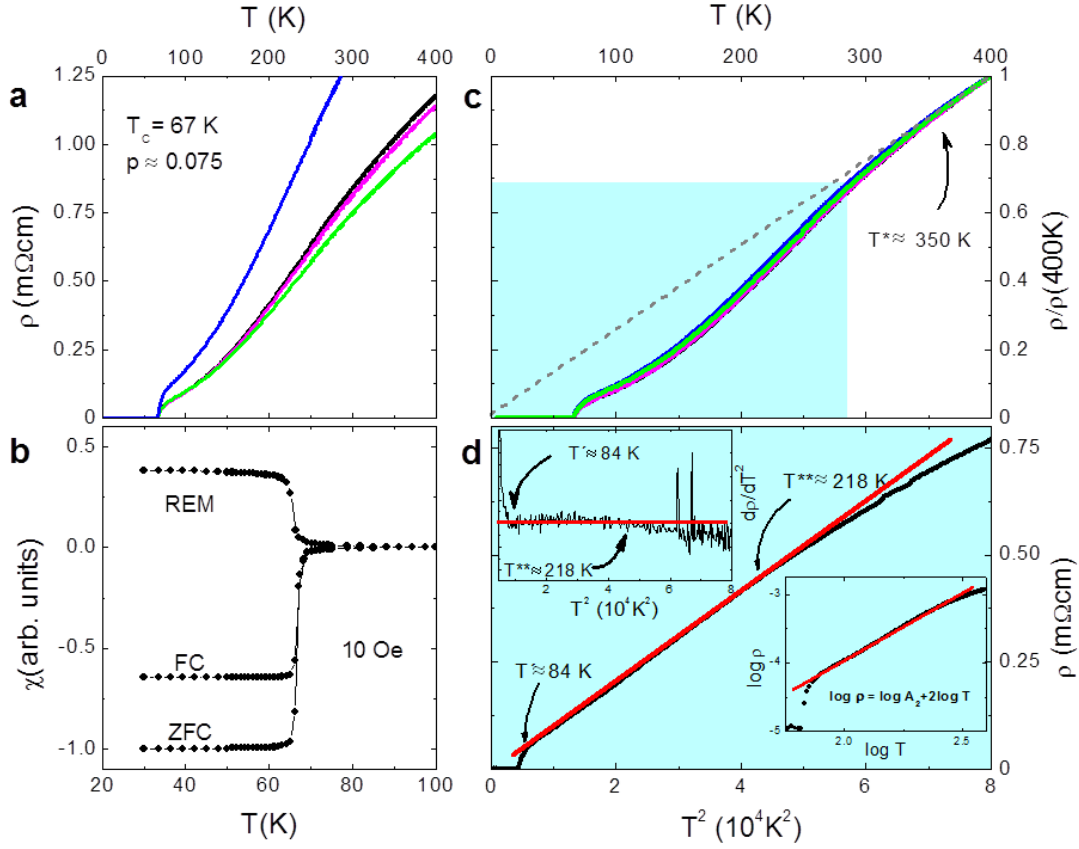

**Figure S1: Resistivity and magnetic susceptibility for underdoped Hg1201 ($T_c$ = 67 K).** (a) After annealing and contact curing (both performed in 0.1 Torr of air at a temperature of 460 °C), the *ab*-plane dc-resistivity was measured for four crystals. (b) Representative *c*-axis magnetic susceptibility data: zero-field-cooled (ZFC), field-cooled (FC) and remnant moment (REM). The ZFC/FC ratio is 65% and the REM is step-like, indicating good bulk sample quality and homogeneous oxygen distribution. (c) Data from a), normalized at the highest temperature (400 K), demonstrating data collapse (as in Fig. 2 for $T_c$ = 80 K). $T^* \approx 350$ K is estimated from the high-temperature linear behavior ($\rho = \rho_0 + A_1 T$ with $\rho_0 \approx 0$), as indicated by the dashed line. (d) Representative resistivity data for one sample at lower temperature (corresponding to the blue region in b)) and d$\rho$/d($T^2$) (upper inset) vs. $T^2$. Quadratic dependence, observed between $T' = 84$ K and $T^{**} = 218$ K, is also apparent from the log($\rho$) vs. log($T$) plot (lower inset).

$T'$ and $T^{**}$ demarcate the deviations from the quadratic behavior of the dc resistivity. To estimate these temperatures for Hg1201, four different approaches were compared: (1) Visual inspection of the $\rho$ vs. $T^2$ curves (Figs. 2, 3 and S1); (2) Greater than 1% deviation of a parabolic fit from the data. A similar approach was also used to estimate $T^*$ as the deviation from linear behavior; (3) The derivative d$\rho(T)$/d$T$ was calculated, and those data that fall on a line with zero intercept correspond to the pure $T^2$ regime; (4) The second derivative was calculated, as shown in insets of Figs. 2d and S1d, and the characteristic temperatures were estimated from the deviation from a constant value larger than the error. All four approaches were found to result in the same characteristic temperatures within 10 K. For LSCO and YBCO, most of the corresponding characteristic temperatures were reported in the literature. When this was not the case, the reported data were digitized and the above approach was followed.

**(5) Determination of coefficient $A_2$ for LSCO ($p = 0.01$)**

The observation of a Fermi-liquid-like resistivity in the lightly-doped cuprates is surprising and was previously reported for LSCO at $p = 0.02$ (8). In Fig. S2, we demonstrate that this is the case even at $p = 0.01$ for prior data for LSCO (9).

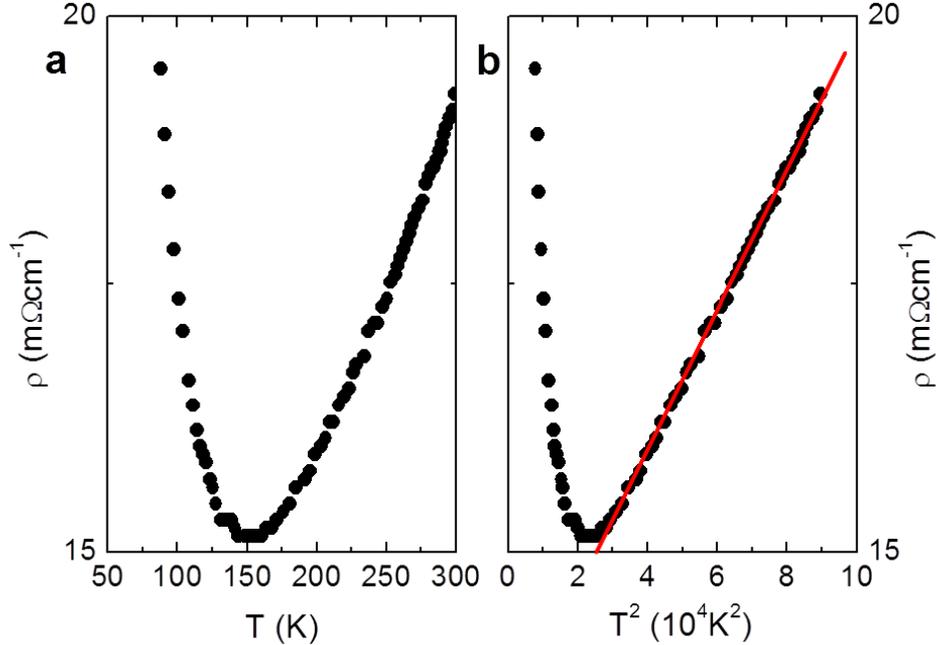

**Figure S2: Temperature dependence of the resistivity for very lightly doped LSCO ($p = 0.01$).** (a) Resistivity up to room temperature, as reported in Ref. (9). (b) By plotting the resistivity versus $T^2$, the quadratic regime is clearly revealed above ~ 170 K. The value $A_2 = 664$ $n\Omega cm^{-1}K^{-2}$ is determined from a fit to the data at higher temperature to a parabolic dependence (red line). As noted before, the metallic behavior is insensitive to the onset of long-range antiferromagnetic order at $T_N = 240$ K (determined from a magnetization measurement), and the low-temperature upturn is due to the onset of localization (9).

**(6) Phase diagrams of YBCO, LSCO and Bi2201**

We show here that the main features of the phase diagrams of single-layer Hg1201 and double-layer YBCO discussed in the main text (Fig. 4) are also present in the structurally more complex lower-$T_c$ compounds LSCO and Bi2201. We are not aware of planar resistivity data that reveal the quadratic resistive regime in Bi2201. For LSCO, the underlying universality is already apparent from the scaling of $A_{1\square}(p)$ and $A_{2\square}(p)$ demonstrated in Fig. 5. The observed universality leads to the conclusion that the Fermi surfaces of the cuprates must be essentially identical near the nodes. Furthermore, it gives insight into why the well-known empirical Presland-Tallon (PT) (57) formula and Obertelli-Tallon-Cooper (OTC) relation (58) hold so well for different cuprate families; PT relate $p$ to $T_c$, whereas OTC relate $p$ to the TEP at 290 K.

For Hg1201, $p$ is taken from the $T_c(p)$ relationship obtained in Ref. ($22$) (data available up to $p = 0.21$). For the previously measured compounds YBCO, LSCO, Bi2201 and Tl2201, we use the published hole concentration estimates. In all cases, either the PT formula or the OTC relation was used to determine the hole concentration. The superconducting dome of Hg1201 (thick blue line in Fig. 4) is extended on the overdoped side using the PT formula. A different method to determine the hole concentration was suggested by Honma and Hor ($52$). However, if correct, this will not affect our main conclusions since $A_{2\square}$ decreases as $1/p$ and the main difference compared to the PT and OTC estimates is on overdoped side of the phase diagram where $p$ is already large.

We use the contour plots of Ref. ($20$) as templates for the phase diagrams of YBCO (Figs. 4 and S3), LSCO (Fig. S4) and Bi2201 (Fig. S5), since they capture the various characteristic temperatures remarkably well. The contour plots were obtained by normalizing $ab$-plane dc-resistivity data at each doping level at high temperature, calculating the second derivative, and linearly interpolating the results at different doping levels. We find that $T^*(p)$ universally falls in the middle of the 'blue' region (which differs from the observations made in Ref. ($20$)), whereas the upper and lower boundaries of the red region in the underdoped regime correspond to $T^{**}$ and $T'$, respectively.

Before considering the phase diagrams of the lower-$T_c$ materials LSCO and Bi2201, we first add in Fig. S3 several experimental details to the phase diagram of YBCO (Fig. 4). The boundary of the antiferromagnetic phase was determined from zero-field μSR measurements ($59$). The Kerr effect indicates broken symmetry in YBCO below $T^{**}$ ($25$). At lower doping and at temperatures slightly below $T^{**}$, magnetic neutron scattering experiments for YBCO provide indirect evidence for an 'electronic liquid crystal' (ELC) state, which was found to be affected by a magnetic field ($60$). It is not clear as yet if the ELC state is a universal property of the cuprates, or a phenomenon specific to YBCO. In the same temperature range (above $p = 0.05$) a feature has been found to appear around 450 cm$^{-1}$ in the $c$-axis optical conductivity of YBCO ($61$). Although this feature was attributed to the Josephson plasmon, it is tempting to relate it to the onset of the ELC state due to its similar doping and temperature dependence. The sharp decrease of the characteristic temperature observed by optical conductivity at lower doping ($61$) coincides with the disappearance of superconductivity at $p = 0.05$ and with a change in the symmetry of the Raman response from $B_{2g}$ to $B_{1g}$ ($62$). Finally, $T'$ attributed to the onset of superconducting fluctuations, as deduced from microwave conductivity ($49$, $50$), falls exactly onto the lower boundary of the red surface in the contour plot, which in turn closely tracks $T_c(p)$. A similarly narrow fluctuation regime that tracks the superconducting dome and behaves rather classically was deduced from dc-resistivity measurements in high magnetic fields (63).

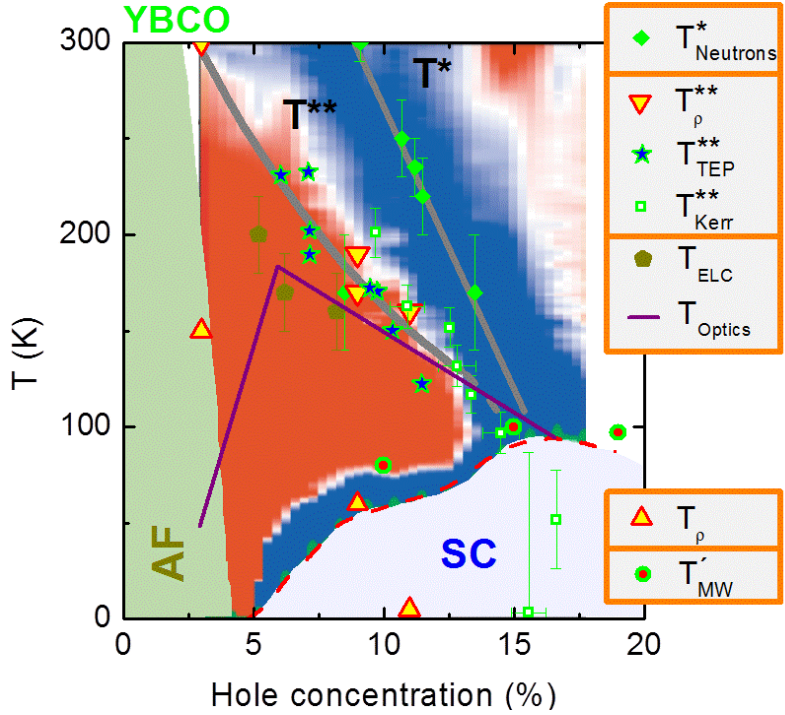

**Figure S3: Phase diagram of YBCO with emphasis on experimental observations related to T\*\*.** The pseudogap temperature $T^*$ lies in the middle of the blue region of the contour plot of Ref. (*20*) and extrapolates to zero temperature at $p \approx 0.19$ (upper gray line). $T^*$ signifies a phase transition to a novel state of matter characterized by an unusual magnetic order that preserves the lattice translational symmetry (*21*). At $p = 0.085$, the ordering temperature was found to sharply decrease (*64*), to a value consistent with $T^{**}$ rather than $T^*$. The upper boundary of the red region corresponds to $T^{**}$. Within error, $T^{**}$ agrees with characteristic temperatures from TEP (*23, 24*), *ab*-plane dc resistivity (in zero (*8, 10*) and high (*11, 12*) magnetic fields), and Kerr effect (*25*). The latter result suggests that, at least in YBCO, $T^{**}$ is therefore associated with a second phase transition above $T_c$. Although it has been suggested that the pseudogap opens below $T^*$, one can also envisage a scenario in which the novel magnetic order at $T^*$ does not gap the Fermi surface and the pseudogap develops as a result of broken crystal translational symmetry at $T^{**}$. Below the slightly lower temperature $T_{ELC}$, evidence for an 'electronic liquid crystal' state has been found. This state appears to be further stabilized by a magnetic field (*60*). Above $p \approx 0.05$, coincident with the onset of superconductivity and a change in the symmetry of the Raman response (*62*), a feature appears around 450 cm$^{-1}$ in the *c*-axis optical conductivity below $T^{**}$ (*61*). The lower bound of the red area closely tracks the superconducting dome and corresponds to the onset of the superconducting fluctuations as determined by microwave conductivity (*50*). The upward triangles mark the lower limit of the $T^2$ behavior in the dc-resistivity in zero ($p = 0.03$) (*8*) and high magnetic fields ($p = 0.09, 0.11$) (*11, 12*). For $p = 0.09$, the quadratic behavior was not observed to very low temperature, presumably because the applied magnetic field was not strong enough to fully suppress superconducting fluctuations. The downward triangles mark the upper onset of the observed $T^2$ behavior.

The phase diagrams of LSCO and Bi2201 are even more complex than that of YBCO. LSCO and Bi2201 are single-CuO$_2$-layer compounds that feature lower structural symmetry, more disorder and significantly lower values of $T_c$ at optimal doping than simple tetragonal Hg1201 (*4*). Nevertheless, the characteristic temperatures (*T\**, *T\*\**, *T'* and *T*c) appear in a very similar fashion on the respective contour plots of Ref. (*20*).

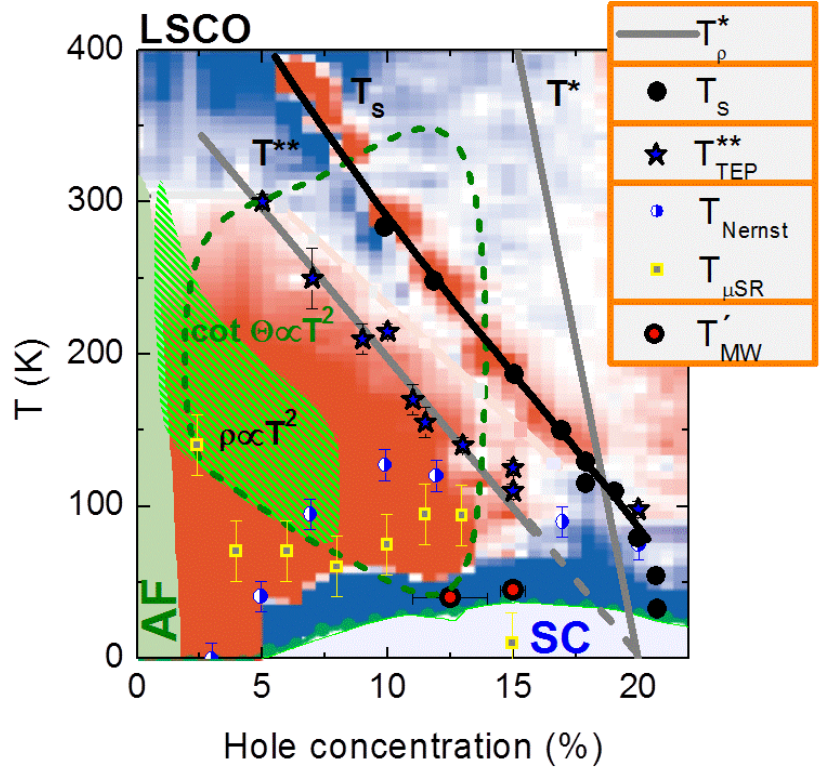

**Figure S4: Phase diagram of LSCO.** *T\** determined from dc-resistivity (*68*) falls in the middle of the blue/white region of the contour plot of Ref. (*20*). The well-known structural transition at *T*S (*69*) modifies the transport properties in its vicinity and is clearly visible. As for YBCO, the upper boundary of the large red area corresponds to *T\*\** as deduced from the peak in the TEP (*65*, *66*). The temperature (*T*Nernst), below which Nernst signal appears (*67*), approximately coincides at intermediate doping with the characteristic temperature from μSR measurements ascribed to charge localization effects (*70*). Also indicated is the onset of superconducting fluctuations as determined from microwave conductivity (*74*), as well as the regions in which $\rho \propto T^2$ and $\cot\theta_H \propto T^2$ have been observed (*8*). The planar resistivity features a quadratic resistivity at intermediate temperatures even at *p* = 0.01 (*9*) (SI Appendix 5).

LSCO exhibits a particularly high pseudogap temperature (*68*), which falls again into the middle of the blue/white region (Fig. S4). The compound undergoes a structural phase transition at *T*S from body-centered tetragonal at high temperature to orthorhombic at lower temperature (*69*), which is clearly evident from the contour plot. A quadratic temperature dependence of the planar resistivity was previously reported for a relatively narrow doping (*p* = 0.02 - 0.08) and temperature (*T* ≈ 60 - 300 K) range (*8*), much narrower than the red contour, which again is bound from above by *T\*\** as determined from TEP. Interestingly, *T\*\**(*p*) lies parallel to *T*S(*p*). The pure $\rho \propto T^2$ behavior in LSCO is bound at high doping by a region in which a substantial Nernst signal (*67*), localization effects (from μSR (*70*) and dc-resistivity (*11*, *71*)), and (non-universal) low-energy stripe correlations (near *p* ≈ 0.10 - 0.12) (*72*) are observed. Whereas the $\rho \propto T^2$ behavior breaks down in this regime, this is not the case for the result $\cot\theta_H \propto T^2$ ($\cot\theta_H = \rho/R_H$, where $R_H$ is the Hall coefficient), which was found to extend throughout a large part of the phase diagram of Fig. S4 (*8*), crossing $T^{**}$ at high temperature and reaching $T'$ at low temperature. In conventional metals, both quadratic temperature dependences signify Fermi-liquid behavior. Since $\rho$ ($\propto 1/n\tau$) and $R_H$ ($\propto 1/n$) are inversely proportional to the effective carrier density ($n$), any change in the latter quantity is identically reflected in both observables, whereas the derived quantity $\cot\theta_H$ is solely proportional to the scattering rate ($1/\tau$). Therefore, in order to explain the reduced $\rho \propto T^2$ regime of LSCO, it is tempting to look for a scenario in which predominately the density of states is affected and the scattering rate in the near-nodal regions remain unchanged. LSCO exhibits considerable localization effects at low temperature and low hole concentrations, as evidenced by significant values of the residual resistivity (Fig. 3) and by the upturn of $T_{\mu SR}$ for $p < 0.07$ in Fig. S4 (*11, 70, 71*). Finally, we note that microwave-conductivity measurements indicate a narrow superconducting fluctuation regime (*73, 74*), as for Hg1201 (*49*) and YBCO (*50*).

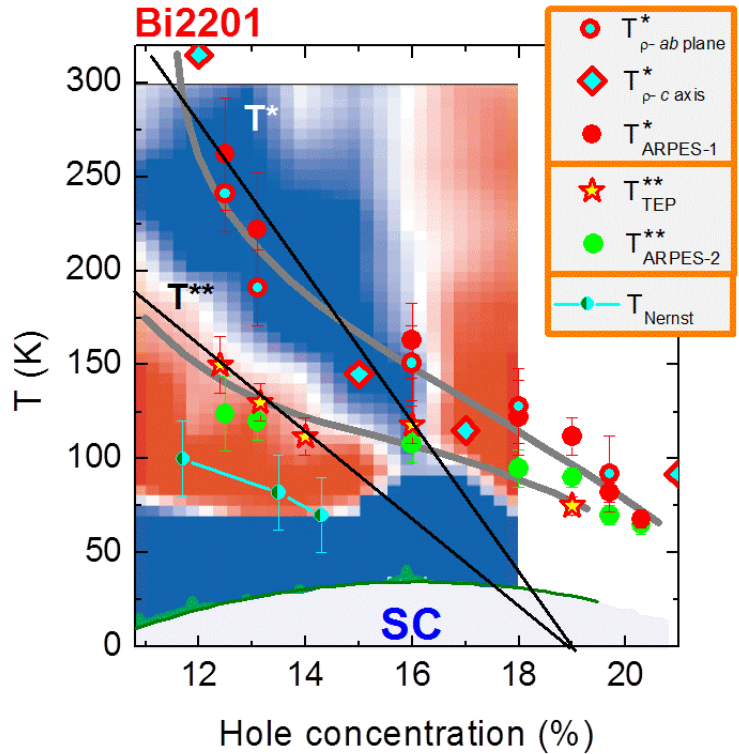

**Figure S5: Phase diagram of Bi2201**. $T^*$ is determined from *ab*-plane (*78*) and *c*-axis (*75*) resistivity as well as photoemission experiments (*78*). $T^{**}$ corresponds to the peak in the TEP (*76, 77*) and to a second temperature observed by photoemission (*78*). Substantial Nernst signal is found to appear at an even lower temperature (*67*). Black and gray lines are guides to the eye.

The above considerations allow us to better understand the phase diagram of Bi2201 (Fig. S5). As for the other compounds, $T^*$ in the underdoped region (deduced from both *ab*-plane and *c*-axis resistivity in this case) falls in the middle of the blue region of the contour plot, and the upper onset of the red region in the underdoped regime corresponds to the peak in the thermoelectric power (*76, 77*). A recent photoemission experiment (*78*) revealed the existence of two distinct temperature scales which agree well with $T^*$ and $T^{**}$. Although in Ref. (*78*) the lower temperature was attributed to the onset of pair formation, other scenarios can be envisaged that involve charge-density-wave formation, nematic order, etc. In more recent photoemission work on (nearly) optimally doped Bi2201(*79*), it was established that the Fermi arc length remains constant between these two characteristic temperatures and then decreases sharply upon cooling below $T^{**}$. (A similar result was obtained for optimally-doped double-layer $Bi_2Sr_2CaCu_2O_{2+\delta}$ (Bi2212)(*79*).This is consistent with the notion that $\rho \propto T^2$ corresponds to a regime of temperature-independent Fermi-arc length and the observation for YBCO (Fig. S3) and LSCO (Fig. S4) that, in zero magnetic field, $\rho \propto T^2$ does not extend into the blue region that surrounds the superconducting dome. It will be important to conduct similar photoemission measurements on underdoped samples of 'cleaner' compounds (e.g., YBCO and Hg1201) for which the underlying quadratic resistive behavior has been clearly established below $T^{**}$ in zero magnetic field.

Given the observation of a Kerr effect in YBCO, the characteristic temperature $T^{**}$ might indeed be universally associated with a (very subtle) phase transition. In fact, the recent observation of a Kerr effect together with supporting evidence from ARPES and nonlinear optical measurements in nearly optimally doped Pb-Bi2201 $(Pb_{0.55}Bi_{1.5}Sr_{1.6}La_{0.4}CuO_{6+\delta})$ seem to support this scenario (*80*). However, near optimal doping, the values of $T^*$ and $T^{**}$ are very similar, making it difficult to distinguish the signatures associated with these two characteristic temperatures.

The red contour for underdoped Bi2201 is relatively narrow when compared to YBCO and LSCO. Its lower edge is matched well by the onset of the Nernst signal, which can be attributed either to the superconducting fluctuations (*67*) or, as discussed above, to the onset of a separate ordering tendency. $T^*(p)$ and $T^{**}(p)$ at lower doping extrapolate linearly to zero around $p = 0.19$ (black lines), as in the other compounds. On the other hand, in the overdoped regime, Bi2201 exhibits a much more gradual doping tendency (grey lines), and both characteristic temperatures remain above the superconducting dome.

An interesting question is why the contour plots of Ref. (*20*) capture so well the various characteristic temperatures of YBCO, LSCO and Bi2201 and, in particular, the underlying quadratic resistive regime. The plots were obtained by normalizing the resistivity data at each doping level by the respective 300 K (400 K in the case of LSCO) value, and by subsequently taking the second derivative of the normalized curves with respect to temperature. The normalization temperatures lie above $T^{**}$ and, in some cases, even above $T^*$. At 300-400 K, the (extrapolated) values of $A_1T$ and $A_2T^2$ are rather similar. Therefore, normalization approximately corresponds to dividing all the curves by $1/p$, since $A_1 \propto A_2 \propto 1/p$. Consequently, in the doping and temperature range where the underlying resistivity is quadratic in temperature, the normalized second derivative is a constant (red areas). For YBCO and LSCO, the lower and upper bounds of these areas in the superconducting doping regime correspond to $T'$ and $T^{**}$, respectively. The structural transition of LSCO is also clearly observed since it introduces an additional scattering mechanism.

Following the above overview of the phase diagrams of hole-doped cuprates, it we briefly comment on their electron-doped counterparts. In recent work on thin films of $La_{2-x}Ce_xCuO_4$ it was suggested that, once the superconductivity is suppressed by either doping (for $x > 0.175$) ($81$) or a magnetic field (48), a low-temperature $\rho \propto T^2$ regime appears. Although it is perhaps tempting to draw parallels between these two rather different systems based on measurements of the resistivity, this could be misleading. In contrast to the hole-doped compounds, the superconductivity in the electron-doped cuprates appears to evolve upon cooling from a regime with a $T$-linear rather than a quadratic scattering rate ($81$). Furthermore, in the electron-doped curpates, the coefficient $A_2$ appears to diverge as the presumed quantum critical point is approached from high dopant concentrations, whereas it remains unchanged (or possibly decreases) in hole-doped cuprates ($14$).

## (7) The overdoped regime

Quantum oscillation ($82$) and photoemission ($83$) experiments for overdoped Tl2201 indicate a large Fermi surface volume, which in accordance with Luttinger's theorem for Fermi Liquids (which holds only in absence of Mott localization precursors) corresponds to $1 + p$ rather than $p$ carriers. Furthermore, for Tl2201 at $p \approx 0.26$, a large magnetic field allowed the suppression of superconductivity, which revealed that the Wiedemann-Franz law in the zero-temperature limit is obeyed, and hence that the fermions that carry heat also carry charge $e$, and that the ground state is therefore indistinguishable from a Fermi-liquid (84). However, the planar resistivity contains a substantial concomitant $T$-linear term down to lowest temperature, a distinctly non-Fermi-liquid property. At $p \approx 0.30$, the low-temperature planar resistivity of Tl2201 still contains a small linear contribution (Figs. 3i and 5), and angle-dependent magnetoresistance measurements have been interpreted as indicative of two distinct scattering rates along the Fermi surface, with distinct temperature dependences ($85$). The appearance of the $T$-linear scattering rate has been linked with the onset of superconductivity with decreasing carrier concentration ($14$, $86$). Moreover, recent Raman scattering experiments for overdoped Tl2201 and double-layer $Bi_2Sr_2(Ca_{1-x}Y_x)Cu_2O_{8+\delta}$ report a well-defined mode with $B_{1g}$ symmetry, consistent with a broken continuous symmetry (with unknown order parameter) at the onset of superconductivity, and it was speculated that this mode is related to the fluctuation spectrum at the origin of superconductivity ($87$). A purely Fermi-liquid-like quadratic resistivity has been observed only for non-superconducting LSCO at $p = x = 0.33$, the highest doping level attained in the cuprates ($16$). Yet the Kadowaki-Woods ratio for LSCO ($p = 0.33$) was found to be significantly larger than for other correlated-electron systems ($16$). Clearly, the evolution toward the putative Fermi-liquid upon exiting the pseudogap state is intricate. Nevertheless, the simple proportionality $A_{2\square} \propto 1/p$ (except near $p*$) shown in Fig. 5 demonstrates an intriguing connection between the underdoped and overdoped (purely $T^2$) regimes of the cuprate phase diagram.

## (8) Multi-band description of the cuprates

Cluster dynamical mean-field calculations of the one-band Hubbard model at intermediate to strong coupling capture several aspects of the experimentally observed doping evolution, such as the nodal/antinodal momentum-space differentiation related to the pseudogap in the antinodal region, which leaves the nodal arc essentially unaffected. (88, 89) In such models, the essentially coherent nodal particles are scattered (36) either by the quantum fluctuations that give rise to the pseudogap formation or through bare Umklapp scattering when the effects of quantum critical fluctuations are negligible, which is certainly the case at very high hole concentrations. The nodal/antinodal dichotomy is further emphasized in the three-band Emery model, where the spectral weight in the nodal regions is associated primarily with planar oxygens, whereas the pseudogap remains confined to the antinodal region (38). Since the Hubbard interaction on oxygen is much smaller than that on copper, the arc is even better protected from magnetic fluctuations than in the single-band model. For the same reason, the nodal particles are also protected from the incoherent copper/oxygen (charge) fluctuations, which generalize the mixed valence fluctuations on the Anderson lattice (90).

## (9) Estimate of energy scale $v_F K_a$ at $p = 0.01$ doping

In the underdoped curpates, the Fermi-liquid-like quadratic resistivity extends to surprisingly high temperatures. We speculate here that this might be the result of a relatively large single-particle energy scale, $v_F K_a$, which in the case of the truncated Fermi surface (arcs) takes on the role of the Fermi energy $E_F$. Here, $K_a$ is the doping-dependent nodal Fermi arc length (30, 33, 91, 92) and $v_F$ the Fermi velocity on the arcs. In a Femi-liquid, $1/\tau \propto T^2$ is attributed to electronic Umklapp scattering. Due to the large value of $v_F \approx 2$ eVÅ determined from photoemission (34, 93), both $k_B T$ and the energy scale associated with effective Umklapp scattering might be small compared to $v_F K_a$, even though $K_a$ decreases with decreasing doping.

We provide a simple estimate of this energy scale for the extreme case of $p = 0.01$. The underlying Fermi surface is approximated by a circle of circumference $2\pi k_F$, with $k_F \approx \pi/a$, where $a \approx 3.9$ Å is the planar lattice constant. We use the photoemission estimates of $K_a$ for Bi2212 listed in Table S1 (92). We assume $T_c = 95$ K at optimal doping and use the PT formula (57) to estimate the hole concentration $p$. Linear extrapolation suggests that ~ 5% of the Fermi surface remains ungapped at $p = 0.01$. The resultant single-particle energy scale $v_F K_a = 2$ eVÅ 0.05 ($2\pi^2/3.9$ Å$^{-1}$) $\approx 0.5$ eV is indeed much larger than $k_B T \sim 0.025$ eV at room temperature.

| $P$ | $T_c$ | $K_a$ (%) |
|---|---|---|
| 0.11 | UD75 | 28 |
| 0.14 | UD92 | 33 |
| 0.20 | OD86 | 44 |

Table S1: Fermi-arc length for underdoped and overdoped Bi2212 (92).

## (10) Estimate of mean-free path

The large scattering rate and the high temperature up to which the quadratic resistivity is observed in the underdoped curpates would seem to suggest that the Fermi-liquid concept has no validity (94, 95). A key requirement for a good metal is that the mean free path $l$ is larger than the lattice constant *(a ~ 4 Å in the case of the cuprates)*. Starting from Ohm´s law,

$$\sigma_{CuO_2} = \frac{1}{\rho_{CuO_2}} = \frac{ne^2 \tau_e(T)}{m} \tag{6}$$

writing $\tau_e(T) = \ell / v_F = m\ell / \hbar k_F$, and assuming a parabolic dispersion in two dimensions ($n = k_F^2 / 2\pi$), the conductivity can be written as

$$\sigma_{CuO_2} = \frac{e^2}{2\pi\hbar}(k_F \ell). \tag{7}$$

The product $k_F\ell$ of the Fermi wave vector and the mean free path is a dimensionless number and turns out to be considerably larger than 1 in the temperature range of interest (up to $\sim 200$ K). This is shown in the inset of Fig. 2a. Using the average Fermi vector of $k_F \approx \pi/a \approx 1$ Å$^{-1}$ associated with the unreconstructed Fermi surface, we estimate the mean free path at $T^{**}$ to be approximately 10 Å, above the Ioffe-Regel limit. Similarly, it was estimated for overdoped Tl2201 (*96*) and LSCO (*16*) that $l \approx 150$ Å below 100 K. Nevertheless, in the $T$-linear regime at higher temperatures, the mean free path eventually becomes smaller than the lattice constant, crossing the Ioffe-Regel limit if the Fermi velocity is assumed to be isotropic. However, the simplest interpretation of our data in the quadratic (and also $T$-linear) regime is that the predominant effect of doping on the resistivity is the reduction of the number of charge carriers ($\rho \propto 1/p$), while the (corresponding) scattering rate is unchanged. If we consider the possibility that the cuprates remain 'nodal' metals even above $T^*$, i.e., that only a fraction (proportional to $p$) of the states at the Fermi level contribute to the planar transport, it follows that even at high temperatures this limit is not crossed. Another important consequence of the above discussion is that the term and the value ($R_Q = h/4e^2 = 6.45$ kΩ) of the quantum resistance, defined based on a simple two-dimensional band, have a questionable meaning in cuprates due to nodal character of these systems.